\documentclass[12pt]{article}
\usepackage{amssymb}
\usepackage{graphicx}
\usepackage{graphics}
\usepackage{epsfig}
\usepackage{setspace,caption}
\makeatletter
\newcommand{\row}[1]{\mathord{\buildrel{\lower3pt\hbox{$\scriptscriptstyle\rightarrow$}}\over #1}}

\newcommand{\dyadic}[1]{\mathord{\dyadic@rrow{#1}}}
\newcommand{\dyadic@rrow}[1]{
\begin{picture}(12,12)(-1,0)
\put(-2,10){\makebox(0,0)[t]{$\scriptscriptstyle\downarrow$}}
\put(-2,11){\makebox(0,0)[l]{$\scriptscriptstyle\longrightarrow$}}
\put(5,0){\makebox(0,0)[b]{$#1$}}
\end{picture}
}
\newcommand{\bra}[1]{\bigl\langle #1 \bigr|}
\newcommand{\ket}[1]{\bigl| #1 \bigr\rangle}
\newcommand{\expect}[1]{\left\langle #1 \right\rangle}

\begin{document}

\begin{center}
\textbf{ Frozen  accelerated information via local operations }
 \vspace{0.5 cm}\\ {\small
N. Metwally\\ \footnote{E-mail: nmetwally@uob.edu.bh} {\small
Department of Mathematics, College of Science, University of
Bahrain, Bahrain.} {\small Department of Mathematics,  Faculty of
Science, Aswan University, Egypt}}

\end{center}

%\maketitle
% \thispagestyle{empty}
% \pagestyle{empty}

%%%%%%%%%%%%%%%%%%%%%%%%%%%%%%%%%%%%%%%%%%%%%%%%%%%%%%%%%%%%%%%%%%%%%%%%%%%%%%%%
\begin{abstract}

In this contribution, we introduce a technique to freeze the
parameters which describe the  accelerated states between two
users to be used in the context of quantum cryptography and
quantum teleportation.  It is assumed that, the two users share
different dimension sizes of particles, where we consider a
qubit-qutrit system.  This technique depends on local operations,
where it is allowed that each particle interacts locally with a
noisy phase channel.  We show that, the possibility of freezing
the information of quantum channel between the users  depends on
the initial state setting parameters, the initial acceleration
parameter strength of the phase channel. It is shown that, one may
increase the possibility of freezing the estimation degree of the
parameters  if only the larger dimension system or both particles
pass through the  noisy phase  channel. Moreover, at small values
of initial acceleration and large values of the channel strength,
the size of  freezing estimation areas increases. The results may
be helpful in the context of quantum teleportation and quantum
coding.
\end{abstract}

%%%%%%%%%%%%%%%%%%%%%%%%%%%%%%%%%%%%%%%%%%%%%%%%%%%%%%%%%%%%%%%%%%%%%%%%%%%%%%%%
\section{INTRODUCTION}
It is well known that, to perform some quantum information tasks
as  teleportation\cite{Bennt}, cryptography \cite{Deutsch},
quantum encoding \cite{Bennt-1,Bose} and  computations
\cite{Nielsen} , one needs to generate  maximum entangled states.
Practically, it is possible to generate these states, but during
their transmission from the source to the users they interact with
their surrounds and consequently  the coherence takes place.
Therefore, these maximum entangled states turn into partially
entangled states and  consequently, their efficiency to perform
the quantum information tasks decrease.  There are different types
of noise which cause this decoherence. One of the noisy channel is
the Unruh effect which is  represented by the acceleration
\cite{Als}. There are several studies  that  have been done to
investigate the behavior of the accelerated systems in different
types of noise.  The possibility of using these accelerated
systems to implement  quantum teleportation is discussed by
Metwally \cite{Metwally2013}. Using the accelerated  system to
perform quantum encoding is investigated in\cite{Metwally2}.
However, some protocols  were introduced to minimize  the losses
of the coherence of the accelerated systems.  Metwally
\cite{Metwally3}  suggested  a protocol  based on weak and reverse
measurement to enhance the local and non-local information of
accelerated two different dimension systems.  Bromely et. al
\cite{Bromely} discussed the possibility of freezing the quantum
coherence of two qubit systems in the presence of different noisy
channels. Recently M.-Ming Du et. al \cite{Ming} investigated the
Unruh effect on the coherence dynamics of accelerated qubit
systems.

As the information of any system is contained in its parameters,
we can minimize the possibility of estimating these parameters or
freezing the degree of estimation during the  transmission process
from one location to another. In this case, if the travelling
states are  captured by adversary, then may he/she get a minimum
information or nothing at all. Moreover, the adversary cannot
dissipate the information and consequently cannot  strayed the
users be sending different information.

Here, we  introduce a technique   to freeze the decoherence due to
the acceleration.  In this context, a system consists of
qubit-qutrit, where it is assumed that only the qubit is
accelerated  with  a uniform acceleration while the qutrit state
in the inertial frame.  In this protocol, we freeze the
information which can be find on the parameters which describe the
accelerated state. The main idea of this technique   is to allow
one or each of these particles (qubit/qutrit) to pass through a
noisy phase channel. We estimate the initial values of the initial
parameter  settings and the strength of the noisy channel  that
maximize,  minimize  and freezing the accelerated state.

The paper is organized as follows. In Sec.II, we describe the
quantum state that is shared between the two users; (Alice) and
(Bob). In Sec. III,the acceleration process is briefly described.
The paper is organized as follows. In Sec.II, we describe the
quantum state that is shared between the  users; Alice and Bob. In
Sec. III, the acceleration process is described briefly. The
possibility of estimating the initial parameters  by using quantum
Fisher information. Finally we discuss our results in Sec. IV.

\section{The suggested quantum system}
\subsection{qubit-qutrit system}
We assume that the  system consists of a qubit(two-levels) system
and a qubit (three-levels system) are initially prepared in an
entangled state.  In the computational basis $\{0,~1\}$, for the
single qubit and  $0, 1, 2$ for the single qutrit, one can write
the  density operator of qubit-qutrit system in the
basis$\ket{00}, \ket{01}, \ket{02}, \ket{10}, \ket{11}$ and
$\ket{12}$ as

 \begin{eqnarray}\label{ini}
 \rho_{ab}&=&\Bigl\{\frac{\mu}{2}(\ket{0}_a\bra{0}+\ket{1}_a\bra{1})\Bigl\}\otimes\ket{1}_b\bra{1}
 %\nonumber\\
+\Bigl\{\frac{\mu}{2}\ket{1}_a\bra{1}+\frac{1-2\mu}{2}\ket{0}_a\bra{0}\Bigr\}\otimes\ket{2}_b\bra{2}
 \nonumber\\
 &+&\Bigl\{\frac{\mu}{2}\ket{0}_a\bra{1}+\frac{1-2\mu}{2}\ket{1}_a\bra{0}\Bigr\}\otimes\ket{0}_b\bra{2}
% \nonumber\\
 +\Bigl\{\frac{\mu}{2}\ket{1}_a\bra{0}+\frac{1-2\mu}{2}\ket{0}_a\bra{1}\Bigr\}\otimes\ket{2}_b\bra{0}
 \nonumber\\
 &+&\Bigl\{\frac{\mu}{2}\ket{0}_a\bra{0}\Bigr\}\otimes\ket{0}_b\bra{0}
 %\nonumber\\
 +
 \Bigl\{\frac{1-2\mu}{2}\ket{1}_a\bra{0}\Bigr\}\otimes\ket{0}_b\bra{1},
\end{eqnarray}
where, $0\leq \mu\leq \frac{1}{2}$ \cite{metwally2016}.
 All the information of this density operator is encoded in the parameter $\mu$, which represents the initial state setting parameter.

\subsection{Acceleration process}
 In this context, we assume that only Alice qubit is accelerated  with  a uniform acceleration $r$, while Bob's qutrit  stays in the inertial.  This means that, Alice qubit will be in Minkowski space,
 Therefore, if the coordinate of a particle is
defined by $(t,z)$ in Minkowski space, then in Rindler space, the
Dirac qubits coordinates  may be defined by $(\tau,x)$, where
\begin{equation}
\tau= a\tanh(t/z),\quad x=\sqrt{z^2-t^2},
\end{equation}
and $\quad -\infty<r<\infty, \quad -\infty<x<\infty$,  $\tan
r_b=exp[-\pi\omega\frac{c}{a_c}]$, ~$0\leq r\leq \pi/4$,
$-\infty\leq a_c\leq\infty$, $\omega$ is the frequency, $c$ is the
speed of  light \cite{Jason2013}. The computational basis
$\ket{0_k}$and $\ket{1_k}$  can be written as
\cite{Friis,Edu,metwally2016},
\begin{eqnarray}\label{trans}
\ket{0_M}&=&\cos r\ket{0_R}_I\ket{0_{R}}_{II}+ \sin
r\ket{1_R}_I\ket{1_{R}}_{II},
 \nonumber\\
 \ket{1_M}&=&\ket{1_R}_I\ket{0_R}_{II}.
\end{eqnarray}

Now, using the transformation (\ref{trans}) and the initial state
(\ref{ini}), the final accelerated state is given by

\begin{eqnarray}\label{Acc-Q}
\rho_{ab}^{ac}&=&\varrho_1\ket{00}\bra{00}+\varrho_2\ket{01}\bra{01}+
\varrho_3\ket{00}\bra{12}
%\nonumber\\
+\varrho_4\ket{12}\bra{00}+\varrho_5\ket{10}\bra{02}+\varrho_6\ket{02}\bra{10}
\nonumber\\
&&+\varrho_7\ket{02}\bra{02}+
\varrho_8\ket{10}\bra{10}+\varrho_9\ket{12}\bra{12}
%\nonumber\\
+\varrho_{10}\ket{11}\bra{11}
\end{eqnarray}
where,
\begin{eqnarray}
\varrho_1&=&\varrho_2=\frac{\mu}{2}\cos^2r,\quad
\varrho_3=\varrho_4=\frac{\mu}{2}\cos r,\quad
%\nonumber\\
\varrho_5=\varrho_6=\frac{1-2\mu}{2}\cos r, \quad
\nonumber\\
\varrho_7&=&\frac{1-2\mu}{2}\cos^2 r,~
\varrho_8=\frac{1-2\mu}{2}+\frac{\mu}{2}\sin^2 r,\quad
\nonumber\\
\varrho_9&=&\frac{\mu}{2}+\frac{1-2\mu}{2}\sin^2 r,
%\nonumber\\
\varrho_{10}=\frac{\mu}{2}(1+\sin^2 r).
\end{eqnarray}

\begin{figure}
  \centering
      \includegraphics[width=0.7\textwidth]{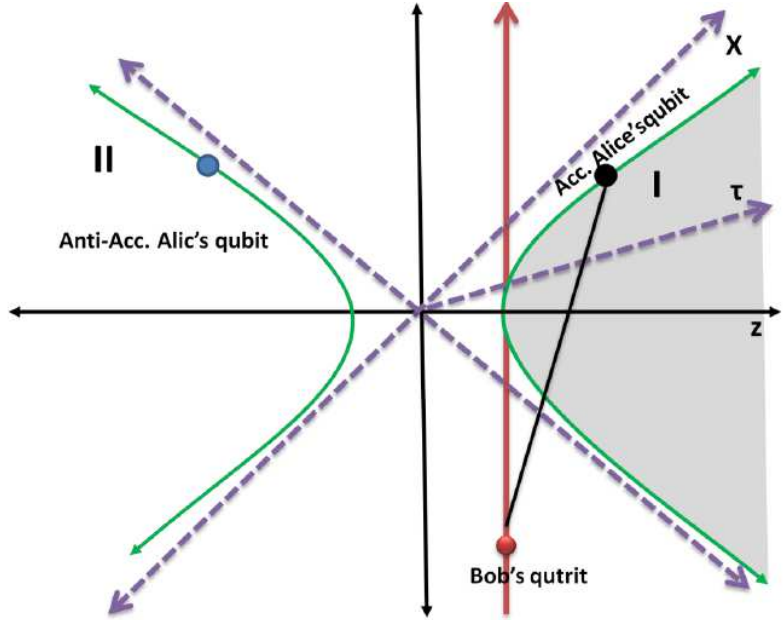}
      \caption{This diagram shows the  dynamics of the accelerated particle in the Rindler space-time.  The accelerated particle travels on a hyperbola in the first region, $I$  with a uniform acceleration,  while its anti-particle travels in the second region, $II$. The non-accelerated particle stays in the inertial frame.}
    \end{figure}

\subsection{Phase channel}
For a single qubit, the phase-flip channel in the computational
basis  set $\{0,~1,~2\}$ may be  described by \cite{Jin},
\begin{eqnarray}
\mathcal{K}_1^{a}=\mathcal{P}_1I_{2\times 2}\otimes I_{3\times 3},
\quad \mathcal{K}_2^{a}=\mathcal{P}_2\sigma_z\otimes I_{3\otimes
3}
\end{eqnarray}
where $\mathcal{P}_1^2+\mathcal{P}_2^2=1$ and $I_{2\otimes 2},
I_{3\otimes 3}$ are the units operators of the qubit and the
qutrit, respectively, y
$\mathcal{P}_1=\sqrt{1-\frac{\gamma_a}{2}}$ and $\gamma_a$ is the
strength of the phase channel. For the single qutrit system the
phase-flip channel is given in the Kraus representation as
\begin{eqnarray}\label{Phasequtrit}
\mathcal{K}_{1}^{b}&=&\mathcal{R}_1I_{2\times 2}\otimes I_{3\times
3}
\nonumber\\
\mathcal{K}_2^b&=&\mathcal{R}_2I_{2\times
2}\otimes\Bigl(\ket{0}\bra{0}+e^{-2i\pi/3}\ket{1}\bra{1}+e^{2i\pi/3}\ket{2}\bra{2}\Bigr),\quad
\nonumber\\
\mathcal{K}_3^b&=&\left(\mathcal{K}_2^b\right)^\dagger
\end{eqnarray}
where $\mathcal{R}_1=\sqrt{1-\frac{2\gamma_b}{3}}$,
$\mathcal{R}_2=1-\mathcal{R}_1^2$, and $\gamma_b$ is the strength
of the channel with respect to the qutrit. Now, we have different
possibility, either the qubit or the qutrit passes through the
phase channel or  both of them are forced to pass through these
noisy channels.

Now, let assume that the accelerated qubit-qutrit system
(\ref{Acc-Q}) passes through the phase-flip channel. The output
state is given by
\begin{eqnarray}\label{Final-phase}
\rho_F&=&\sum_{J=1}^3\sum_{i=1}^2\left\{\mathcal{R}^b_j\mathcal{K}^a_i\rho_{ab}\Bigl({\mathcal{K}_i^a}\Bigr)^\dagger\Bigl({\mathcal{R}_j^b}\Bigr)^{\dagger}\right\}
\end{eqnarray}
In the computational basis the final state (\ref{Final-phase}) is
given by

\begin{eqnarray}
\rho_F&=&\varrho_1\ket{00}\bra{00}+\varrho_2\ket{01}\bra{01}+
\varrho_7\ket{02}\bra{02}
+\varrho_8\ket{10}\bra{10}+\varrho_{10}\ket{11}\bra{11}
\nonumber\\
&+&\varrho_9\ket{12}\bra{12}
+(1-\gamma_a)\eta_b\Bigl(\varrho_3\ket{00}\bra{12}+\varrho_4\ket{12}\ket{00}
%\nonumber\\
+\varrho_5\ket{10}\bra{02}+\varrho_6\ket{02}\bra{10}\Bigr)
\end{eqnarray}
where $\eta_b=1-\frac{2\gamma_b}{3}(1-\cos(2\pi/3))$.

The main task now is estimating the initial parameter of the state
settings, $(\mu)$  and of the initial strength  channels, $\gamma$
, where we set  $\gamma_a=\gamma_b=\gamma$.

\section{Fisher Information}
\subsection{Mathematical Form}

Quantum Fisher information represents a central role in the
estimation theory, where it can be used  to quantify some
parameters that cannot be quantified directly \cite{Yue}. In this
subsection, we review  the    mathematical form of Fisher
information.  Let $\eta$ be the parameter to be estimated. The
spectral decomposition of the density operator is given by
$\rho_{\eta}=\sum_{j=1}^n\kappa_j\ket{\psi_j}\bra{\psi_j}$, where
$\kappa_j$and $\ket{\psi_j}$ are the eigenvalues and the
corresponding eigenvectors of the state $\rho_\eta$. The Fisher
information with respect to the parameter $\eta$ is defined as
\cite{Helstrom}
\begin{equation}
\mathcal{F}_{\eta}=tr\{\rho_{\eta}L^2_\eta\}, ~
\end{equation}
where the  symmetric logarithmic derivative $L_\eta$ is a solution
to the  equation
$\frac{\partial\rho_{\eta}}{\partial\eta}=\frac{1}{2}(\rho_{\eta}L_{\eta}+L_{\kappa}\rho_{\kappa})$.
 Using the spectral decomposition and the identity
$\sum_{j=1}^n\ket{\psi_j}\bra{\psi_j}=1$ in Eq.(5), one can obtain
the final form of the Fisher information as\cite{Xiao,Yao,Wei},
\begin{equation}
\mathcal{F}=\mathcal{F}_{cl}+\mathcal{F}_{qu}-\mathcal{F}_{mix},
\end{equation}
where
\begin{eqnarray}\label{Fisher}
\mathcal{F}_{cl}&=&\sum_{j=1}^{n}\frac{1}{\kappa_j}\left(\frac{\partial
\kappa_j}{\partial\eta}\right)^2,
\nonumber\\
\mathcal{F}_{qu}&=& 4\sum_{j=1}^{n}\kappa_j\Bigl(
\expect{\frac{\partial\psi_j}{\partial\eta}\Big|\frac{\partial\psi_j}{\partial\eta}}
-\Bigl|\expect{\psi_j\Big|\frac{\partial\psi_j}{\partial\eta}}\Bigr|^2\Bigr),
\nonumber\\
\mathcal{F}_{mix}&=&8\sum_{j\neq\ell}^{n}\frac{\kappa_j\kappa_\ell}{\kappa_j+\kappa_\ell}
\Big|\expect{\psi_j|\frac{\partial\psi_\ell}{\partial\eta}}\Big|^2.
\end{eqnarray}
 The summations on the first and the third terms over all
$\kappa_j\neq 0$ and $\kappa_j+\kappa_\ell\neq 0$. The first and
the second terms represent the classical Fisher information,
$(\mathcal{F}_c)$, and   the  quantum Fisher information of all
pure states, $(\mathcal{F}_p)$, respectively. The third term,
$(\mathcal{F}_m)$ is stemmed from the mixture of the pure states.
It is clear that, Fisher information for a pure state is just the
first two terms, namely
$\mathcal{F}_{pure}=\mathcal{F}_c+\mathcal{F}_p$, while for a
mixed state the third term is subtracted. Therefore, the quantum
Fisher of a pure state is larger than that displayed for a mixed
state \cite{Xiao,Yao,Wei}.
\begin{figure}
  \centering
     \includegraphics[width=0.4\textwidth]{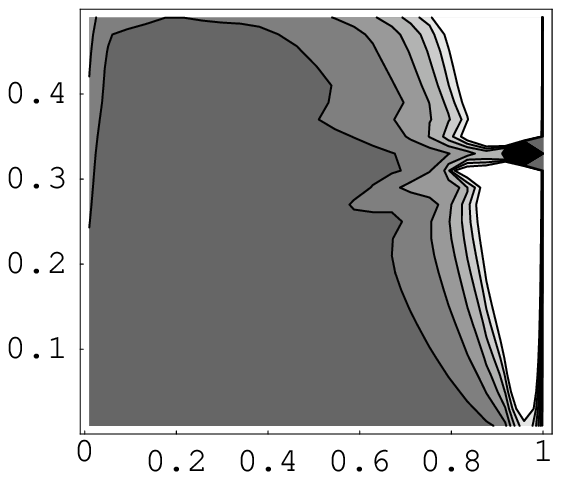}
     \put(-80,-2){\large $\gamma$}
     \put(-185,90){\large $\mu$}
     \put(-185,155){$(a)$}~~\quad
\includegraphics[width=0.4\textwidth]{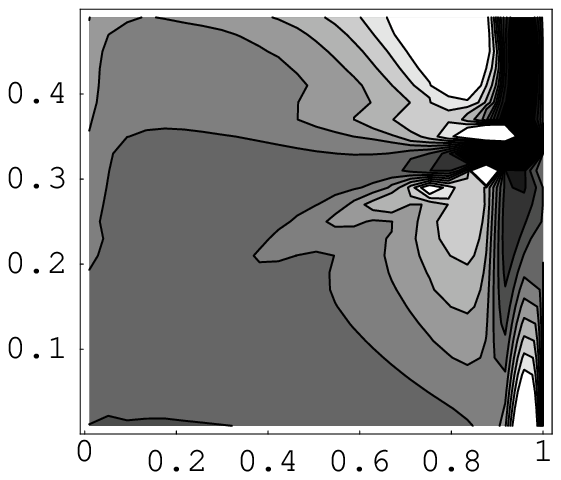}
\put(-80,-2){\large $\gamma$}
     \put(-185,90){\large $\mu$}
     \put(-185,155){$(b)$}
           \caption{Fisher information with respect to the acceleration parameter, $\mathcal{F}_r$ where the initial acceleration (a) $r=0.1$ and
           (b)$r=0.2$.}
    \end{figure}

\begin{figure}
  \centering
     \includegraphics[width=0.6\textwidth]{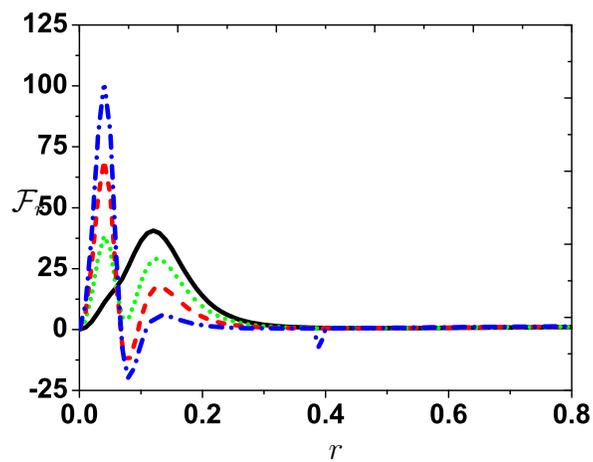}
      \put(-130,6){$r$}
     \put(-250,100){$\mathcal{F}_r$}
           \caption{Fisher information, $\mathcal{F}_r$ for an accelerated system initially prepared
            $\gamma=0.99$ and $\mu=0.01,0.1,0.2,3$ for the solid, dot, dash and  dash-dot, respectively.}
    \end{figure}

\subsection{Only the qubit passes through the phase channel}
Fig.(2a) displays the behavior of the quantum Fisher information
$\mathcal{F}_r$ for any initial state settings and initial channel
strength $\gamma$ at a small  initial value of the acceleration
$(r=0.1)$.  The degree of the darkness displays the possibility of
estimating the parameter $r$. However, as the brightness increases
the estimation's degree of the parameter $r$ increases. Different
regions show that, the Fisher information $\mathcal{F}_r$ is
frozen.   Fig.(2b)   displays  the behavior of $\mathcal{F}_r$ at
larger values of the initial acceleration where we set $r=0.2$. In
general, the behavior is similar to  that displayed in Fig.(2a).
However, the size  of the estimation  areas are changed and
irregularity takes place.

From Fig.(2), for  small values of the initial acceleration, the
sizes of the frozen areas are wider than those displayed at larger
values of initial acceleration.

In Fig.(3), we show the effect of   maximum value of channel
strength on $\mathcal{F}_r$, where we set $\gamma_a=0.99$. At zero
acceleration, the Fisher information increases suddenly for  any
initial state setting to reach its  maximum values.  These maximum
values depend on the initial state settings, where they are higher
at larger values of $\mu$. However, at smaller values of $r$, the
sudden decay phenomena is predicted at larger values of $\mu$.
Moreover, the behavior of the  quantum Fisher information,
$\mathcal{F}_r$ is almost frozen as one increases the
acceleration.

\begin{figure}[t!]
  \centering
\includegraphics[width=0.4\textwidth]{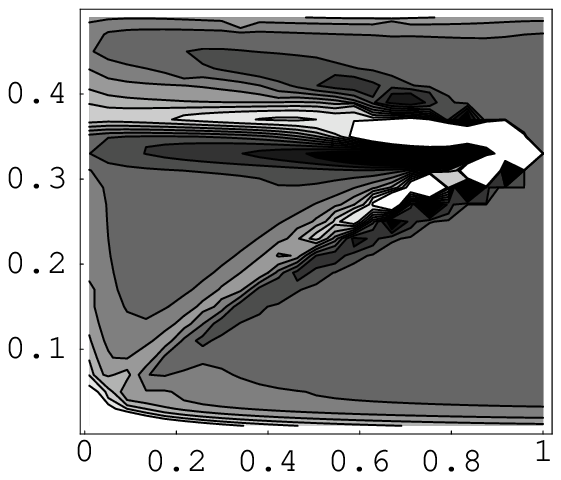}
    \put(-80,-2){\large $\gamma$}
     \put(-185,90){\large $\mu$}
      \put(-185,155){$(a)$}~~\quad
\includegraphics[width=0.55\textwidth]{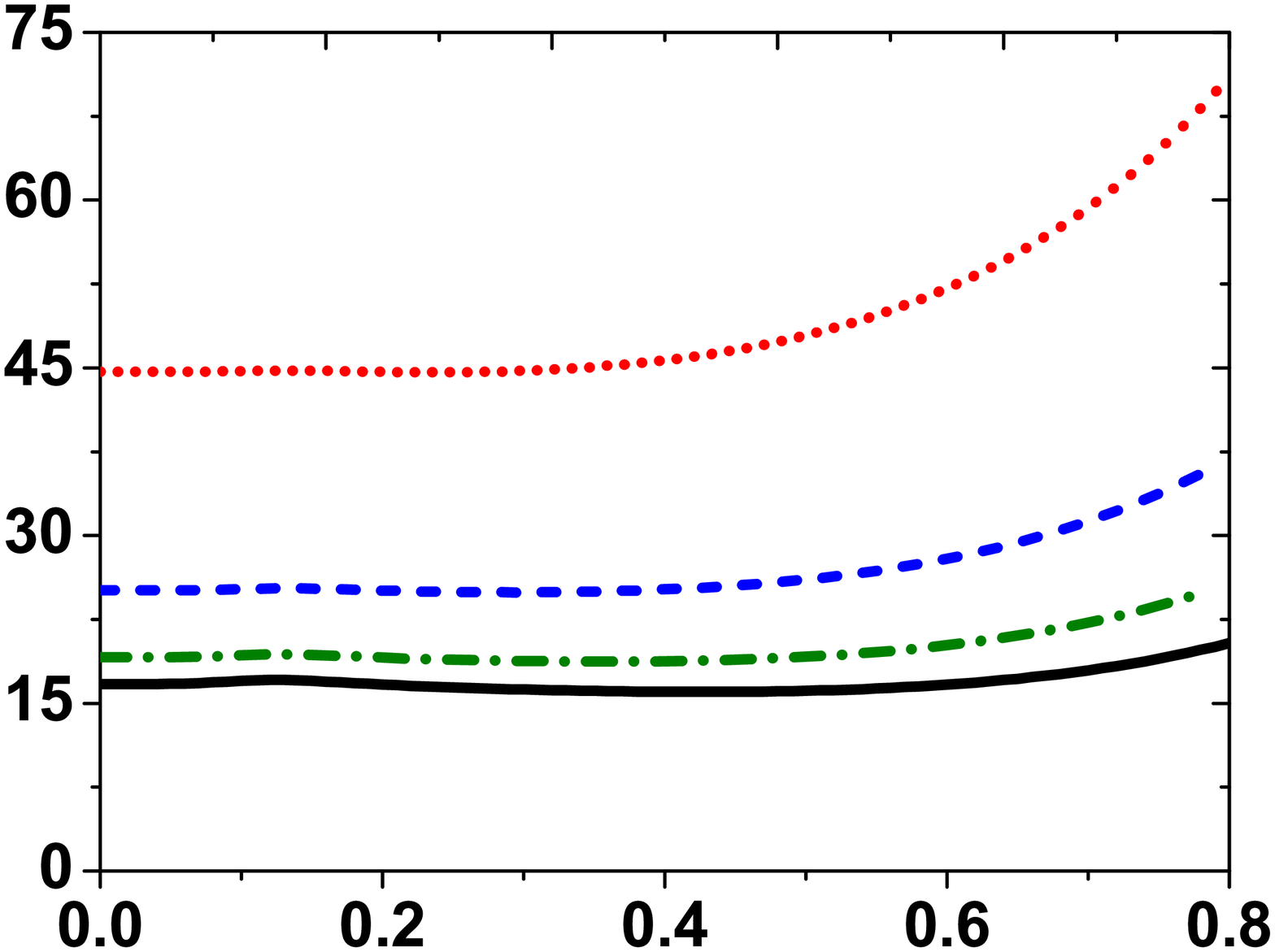}
 \put(-115,6){$r$}
     \put(-240,90){$\mathcal{F}_\mu$}
      \put(-240,155){$(b)$}
      \caption{ (a)The Fisher information with respect to the initial state setting parameter $\mu$  at an initial acceleration $r=0.2$.
      (b)Fisher information $\mathcal{F}\mu$, where $\gamma=0.99$
               and    $\mu=0.05,0.1,0.15,0.2$  for the dot, dash, dash-dot and solid, respectively}
    \end{figure}

Fig(4a) shows the  values of $\mu$ and$\gamma$ that freeze the
quantum Fisher information with respect to the initial state
setting parameter $\mu$. Different degrees of darkness  are
displayed. This means that,  in  these regions the quantum Fisher
information is frozen. The degree of brightness increases at
larger values of $\mu$ and $\gamma$. Fig.(4b) shows the frozen
effect of the channel and the gradual increasing of
$\mathcal{F}_\mu$ as the acceleration increases. However, the
increasing rate of $\mathcal{F}_\mu$ increases for smaller values
of the initial state parameter $\mu$ and larger values of $r$.

\subsection{Only the Qutrit  passes through the phase channel}.
Figs.(5)   display the areas in which one may estimate the
acceleration parameter by means of the quantum Fisher information
$\mathcal{F}_r$. It is clear that, the degree of brightness
increases as $\mu$  and $r$ increase.  The similar color  degree
means that in  these regions the Fisher information is Frozen .
The size of the regions depend on the initial acceleration, e.g.
at $r=0.5$ the size areas are smaller than that displayed at
$r=0.1$. Also, at  large values of $\mu$ or $r$, the possibility
of estimating the acceleration parameter $r$ increases.

\begin{figure}
  \centering
\includegraphics[width=0.4\textwidth]{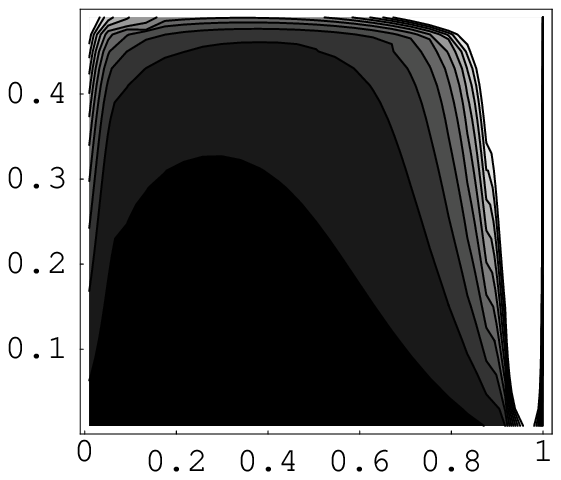}
      \put(-80,-2){\large $\gamma$}
     \put(-185,90){\large $\mu$}
      \put(-185,155){$(a)$}~~\quad
     \includegraphics[width=0.4\textwidth]{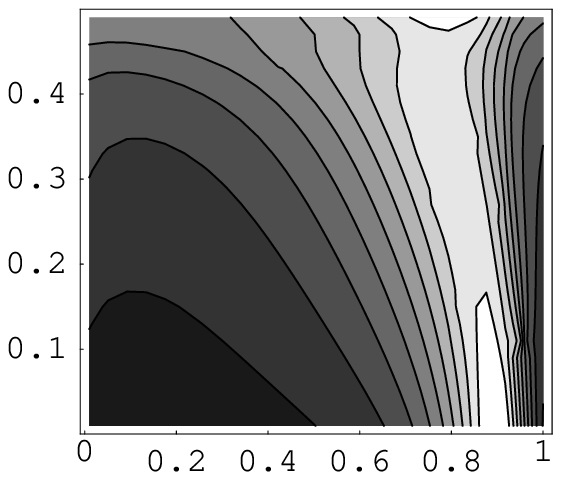}
      \put(-80,-2){\large $\gamma$}
     \put(-185,90){\large $\mu$}
      \put(-185,155){$(b)$}
      \caption{ The Fisher information with respect to the acceleration parameter when only the qutrit passes through the phase channel with(a)  $r=0.1$.
      and (b) $r=0.5$}
    \end{figure}

\begin{figure}
  \centering
     \includegraphics[width=0.55\textwidth]{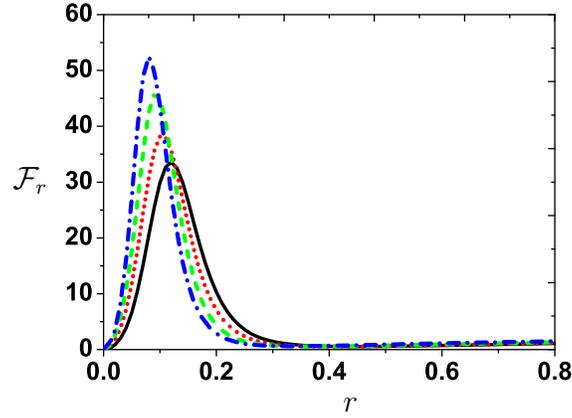}
       \put(-115,6){$r$}
     \put(-240,90) {$\mathcal{F}_r$}
           \caption{$\mathcal{F}_r$ at $\gamma=0.99$ and  for different values of $\mu=0.1,0.2,0.3,0.4
$ for the solid, dot, dash and dash-dot curves, respectively.}
    \end{figure}
Fig.(6),  shows the effect of the  maximum value of the channel
strength, where we set $\gamma=0.99$. From this figure one sees
that, the possibility of estimating the acceleration parameter,
$r$, increases as the initial  acceleration increases to reach its
maximum value. As $r$ increases, the Fisher information
$\mathcal{F}_r$ decreases gradually and consequently the
estimation  degree of  $r$ decreases as  the  initial state
parameter setting $\mu$  increases. However, the sudden
increasing/decreasing behaviors of $\mathcal{F}_r$ are displayed
for  large values of the noise channel strength. The freezing
effect of the channel strength is depicted  for  value of
$r\in[0.4,0.8]$ and any initial state parameter settings $\mu$.

\begin{figure}
  \centering
\includegraphics[width=0.4\textwidth]{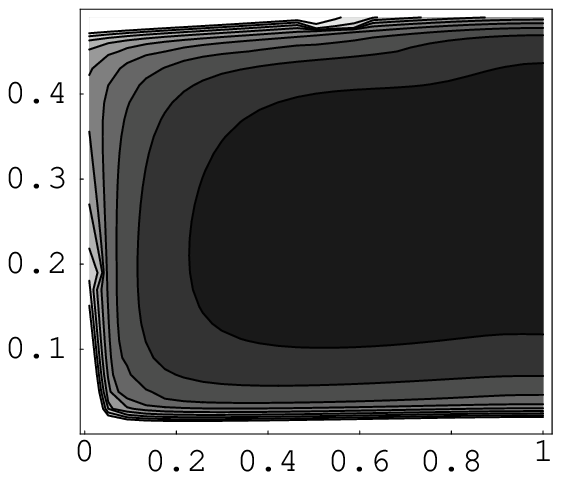}
     \put(-80,-2){\large $\gamma$}
     \put(-185,90){\large $\mu$}
      \put(-185,155){$(a)$}~~\quad
     \includegraphics[width=0.4\textwidth]{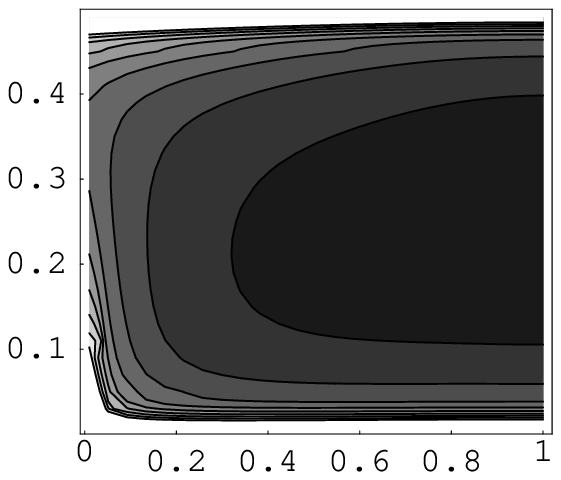}
     \put(-80,-2){\large $\gamma$}
     \put(-185,90){\large $\mu$}
      \put(-185,155){$(b)$}
           \caption{ The Fisher information with respect to the initial state settings $\mathcal{F}\mu$
            when only the qutrit passes through the phase channel with (a)$r=0.5$ and (b) $r=0.8$.}
    \end{figure}

Fig.(7) displays the effect of different  values of the initial
acceleration  on the estimation degree of the parameter $\mu$. It
is clear that, the possibility of estimating $\mu$ increases at
small values of $\gamma$ and larger values of $\mu$ or small
values of $\mu$ and an arbitrary values of the channel' strength
$\gamma$. The  dark areas decrease as $r$ increases, namely, the
possibility of freezing the information  decreases as $r$
increases.

\subsection{The qubit and  the Qutrit  pass through the phase channel}

\begin{figure}[b!]
  \centering

       \includegraphics[width=0.4\textwidth]{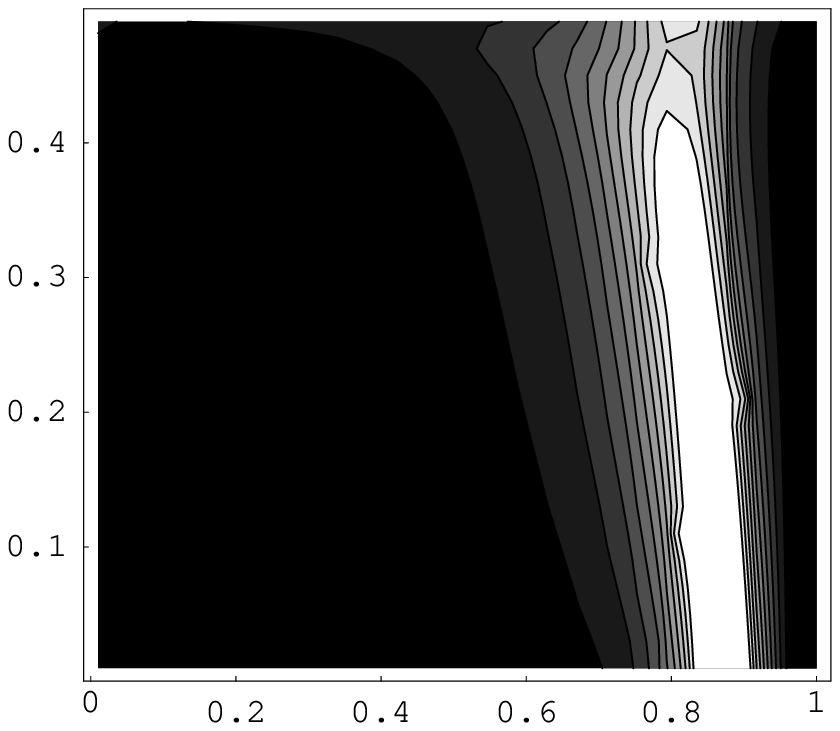}
        \put(-80,-5){$\gamma$}
   \put(-180,80){$\mu$}
     \put(-185,140){$(a)$}~~\quad
\includegraphics[width=0.4\textwidth]{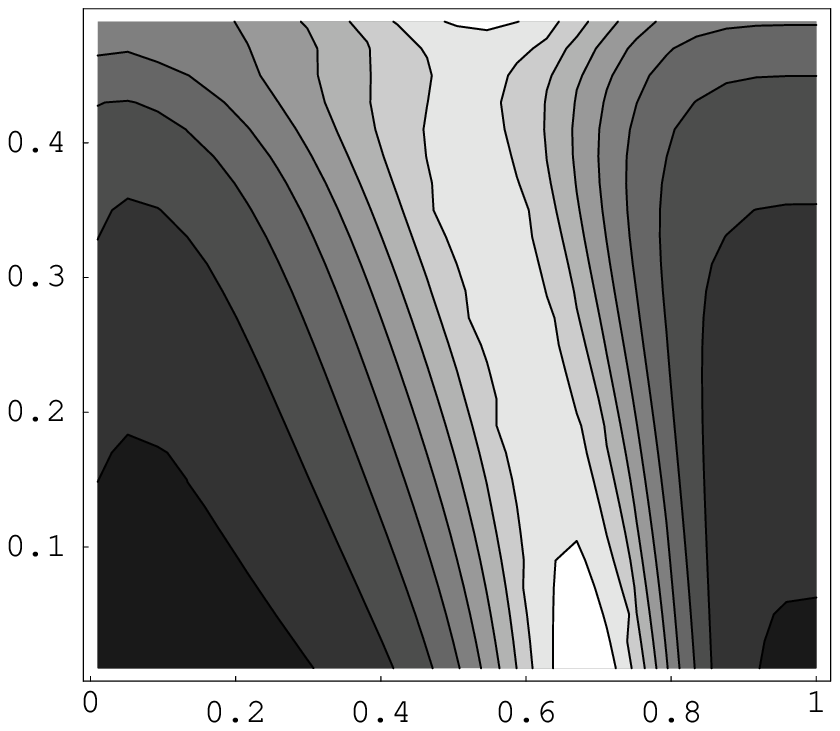}
     \put(-100,-2){\large $\gamma$}
     \put(-210,95){\large $\mu$}
      \put(-185,155){$(b)$}
             \caption{ The Fisher information with respect to the acceleration parameter, $\mathcal{F}_r$ when
              both particles  pass through the phase channel with (a) $r=0.2$ and (b) $r=0.5$.}
    \end{figure}

It is well know that, the Unruh effect causes a coherence of the
accelerated information. Therefor, we investigate the effect of
the initial state settings and the channel' strength at different
values of the acceleration. Fig.(8) describes the behavior of the
Fisher information with respect to the acceleration parameters. In
Fig.(8a), It is assumed that, the qubit  is accelerated with  a
small  acceleration, where we set $r=0.1$.  It is clear that, for
$\mu\leq 6$, and any initial value of channel' strength, the
Fisher information $\mathcal{F}_\mu$ is almost frozen. At larger
values of $\mu>0.6$, the quantum Fisher information increases
gradually to reach its maximum values at $\mu\simeq 0.9$. However,
at further values of the channel strength $\gamma$,
$\mathcal{F}_r$ decreases suddenly to vanish completely at
$\mu\simeq 1$ and the frozen behavior is observed again at
$\mu=1$.

  Fig.(8b), shows  the behavior of
$\mathcal{F}_r$ at larger values of the initial acceleration
parameter. It is clear that as $r$ increases, the range of the
channel' strength in which one can freeze the accelerated
information decreases.  The upper bounds of the quantum Fisher
information $\mathcal{F}_r$ decreases as the  initial acceleration
increases.  The phenomena of the sudden increase/decrease is
displayed for  small values of $r$, while the gradual behavior of
$\mathcal{F}_r$ is depicted for larger values of $r$.

\begin{figure}[b!]
  \centering
\includegraphics[width=0.4\textwidth]{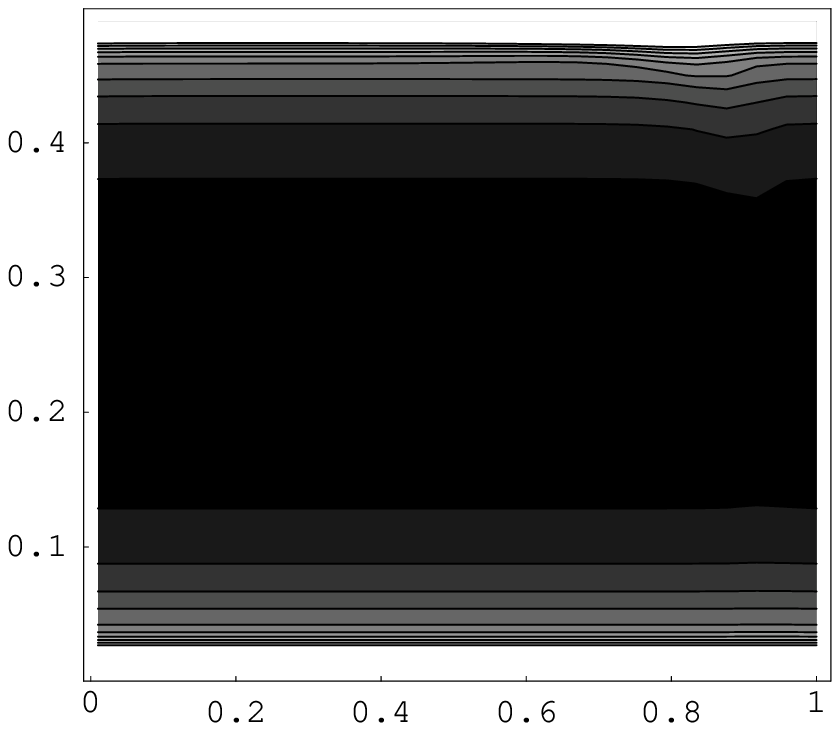}
     \put(-80,-2){\large $\gamma$}
     \put(-185,90){\large $\mu$}
      \put(-185,155){$(a)$}~~\quad
     \includegraphics[width=0.4\textwidth]{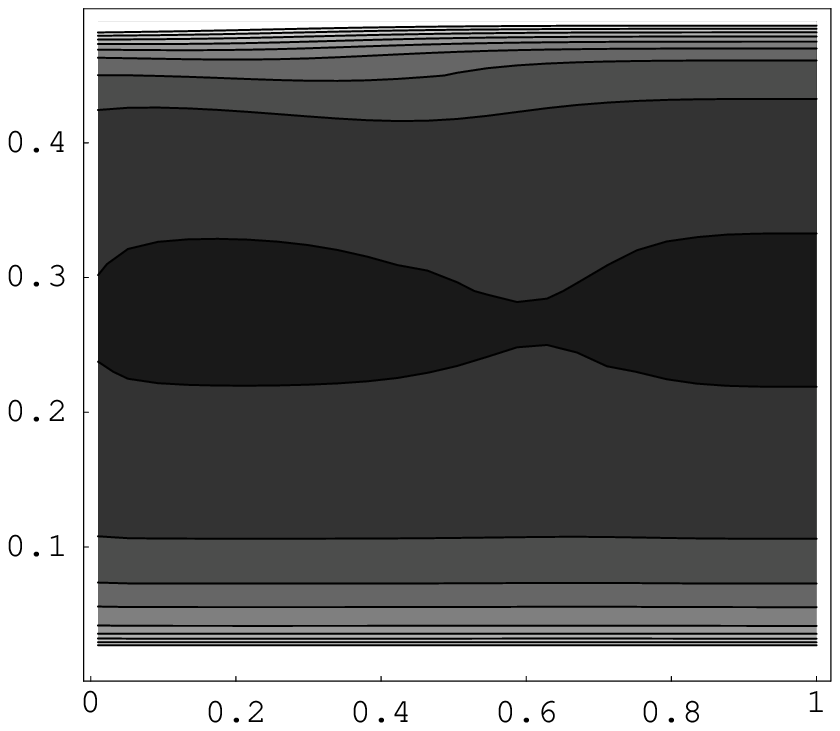}
     \put(-80,-2){\large $\gamma$}
     \put(-185,90){\large $\mu$}
      \put(-185,155){$(b)$}
                \caption{ The Fisher information  $\mathcal{F}_\mu$ when both particle  passe through the phase channel with
                (a) $r=0.1$ and (b) $r=0.5$.}
    \end{figure}

 In  Fig.(9a), the
behavior of the quantum Fisher information $\mathcal{F}_\mu$ is
displayed for small value of the initial acceleration, where we
set $r=0.1$ and arbitrary values of $\mu$ and $\gamma$. It is
clear that, at small values of $\gamma$ and $\mu$ the Fisher
information $\mathcal{F}_\mu$ is maximum. However, as one
increases these initial parameters, $\mathcal{F}_\mu$ decreases
gradually to vanishes completely for $\mu\simeq\in[1.4,3.75]$ and
any arbitrary values of the channel' strength $\gamma$. On the
other hand, the sudden increasing behavior of  $\mathcal{F}_\mu$
is depicted at larger values of the initial state settings, namely
$\mu>0.4$ and larger values of the initial channel' strength
($\gamma>0.8)$.

  The bright regions
indicate that this parameter may be estimated, while it can't be
estimated in the dark regions. Also, the degree of estimation is
almost the same for each  region, namely the  accelerated channel
is frozen.  Further, in the dark region the accelerated state is
not only frozen but also   the estimation degree of $\mu$ is
almost zero.

In Fig.(9b), we show that behavior of $\mathcal{F}\mu$ at larger
values of the initial acceleration, where we set $r=0.5$.
 One  can estimate the initial state setting
parameter $\mu$ in the  bright regions and the degree of
estimation decreases as the darkness of the region increases.  One
can  also pick the  values of the  initial parameters which freeze
the coherence due to the acceleration $r$.

\section{CONCLUSIONS}

In this contribution, we estimate the initial state settings  and
the acceleration that  freeze  the accelerated state.  It is shown
that, the degree of estimation depends on either one or both
particles are affected by the  phase noise channel. If only the
qubit is allowed to pass through the phase channel, then one may
freeze the information contained in the accelerated state  for
smaller values of the acceleration,  arbitrary  initial state
settings and arbitrary values of the channel strength. However, at
larger values of the acceleration, one  may be able to freeze the
accelerated state by increasing the channel' strength. Moreover,
the Fisher information with respect to the initial state setting
parameter may be frozen at smaller values of initial state
settings and larger values of the phase channel strength.

On the other hand, if only the qutrit passes through the phase
channel, the areas in which the Fisher information is frozen  are
more regular and the estimation degree is smaller, where the
degree of darkness are much larger than that displayed  when only
the qubit passes through the phase channel. The frozen areas
decrease as the initial acceleration increases. One may also
freeze  the Fisher information  if the phase channel' strength is
large.

Finally, if  both particles pass through the phase channel, then
at small value of the initial acceleration, the freezing areas are
much wider than those displayed in the previous two cases.
However, for larger  initial acceleration, the possibility of
freezing the accelerated Fisher information decreases, where the
size of the  dark areas decreases. The decreasing rate of theses
areas is larger if one estimate the Fisher information with
respect to the acceleration parameters. However, the size of the
freezing area with respect to the initial state setting parameter
is large.

{\it In conclusion,}  it is possible to minimize or freeze the
estimation degree of the parameters which describe the accelerated
qubit-qutrit system by using quantum Fisher information.
Therefore the travelling state is protect from  any  Eavesdropper
and consequently these state may be useful in context of quantum
teleportation and encoding.

\end{document}